\documentstyle[12pt,epsf]{article}
\newcommand {\be}{\begin{equation}}
\newcommand {\ee}{\end{equation}}
\newcommand {\bea}{\begin{eqnarray}}
\newcommand {\eea}{\end{eqnarray}}
\newcommand {\nono}{\nonumber}
\def\vep{\varepsilon}
\def\mn{\mu\nu}
\def\der{\partial}

\def\la{\lambda}
\def\tF{\tilde{F}}
\def\tV{\tilde{V}}
\def\Tr{\mbox{Tr}}

\begin{document}
\setlength{\oddsidemargin}{0cm}
\setlength{\baselineskip}{7mm}

\begin{titlepage}
 \renewcommand{\thefootnote}{\fnsymbol{footnote}}
$\mbox{ }$
\begin{flushright}
\begin{tabular}{l}
KUNS-1723 \\
hep-th/0106103\\
June 2001
\end{tabular}
\end{flushright}

~~\\
~~\\
~~\\

\vspace*{0cm}
    \begin{Large}
       \vspace{2cm}
       \begin{center}
         {Strings as Flux Tube and Deconfinement on Branes \\
                         in Gauge Theories}
\\
       \end{center}
    \end{Large}

  \vspace{1cm}

\begin{center}

          Hikaru K{\sc awai}\footnote
           {
e-mail address : hkawai@gauge.scphys.kyoto-u.ac.jp}{\sc and}
          Tsunehide K{\sc uroki}\footnote
           {
e-mail address : kuroki@gauge.scphys.kyoto-u.ac.jp}

           {\it Department of Physics, Kyoto University,
Kyoto 606-8502, Japan}\\
\end{center}

\vfill

\begin{abstract}
\noindent
We propose gauge theories in which the unstable branes and the fundamental 
string are realized as classical solutions. While the former are 
represented by domain wall like configurations of a scalar field 
coupled to the gauge field, the latter is by a confined flux tube in the bulk. 
It is shown that the confined flux tube is really a source of the bulk 
$B$-field. Our model also provides a natural scenario of the confinement 
on the brane in the context of the open string tachyon condensation. 
It is also argued that the fundamental string can be
realized as a classical solution in a certain IIB matrix model 
as in our model. 

\end{abstract}

\vfill
\end{titlepage}
\vfil\eject

\section{Introduction}
\setcounter{equation}{0}

It has been believed that string theory can be nonperturbatively
defined in terms of a gauge theory. In fact, candidates
\cite{BFSS,IKKT,maldacena} for the nonperturbative definition 
of string theory are formulated as gauge theories in lower dimensions 
and some evidences have been found that these models really
contain fundamental strings \cite{BS,DVV,FKKT}. Unfortunately, however, 
we still do not know much about how to extract the fundamental string 
degrees of freedom explicitly from these gauge theories. 
{}From this point of view, it must be important to construct 
classical solutions corresponding to fundamental strings in gauge theories. 

On the other hand, physics of the (open string) tachyon condensation 
draws much attention recently. There, it is conjectured that after the
tachyon condensation, unstable D-branes disappear and end up with a
bunch of closed strings. If a gauge theory is supposed to be a
nonperturbative definition of string theory, it is desirable to
describe such phenomena in terms of a gauge theory.    

In light of these situations, it is useful to construct a gauge theory 
which realizes a fundamental string and an unstable brane as 
classical configurations. This is exactly the aim of this paper. 
The idea is the following: let us consider a gauge theory coupled to a
scalar field which has an unstable domain wall solution
--- ``brane''. Suppose in the bulk, the gauge theory is in the
confinement phase, while on the brane, it is in the Coulomb 
(deconfinement) phase. 
Then we have the standard Abelian gauge theory on the brane and 
the confined flux tube in the bulk plays a role of a fundamental string. 
The flux tube attached to the brane will provide a deconfined flux 
on the brane. As the unstable brane decays, this flux tends to be
confined as in the bulk, and finally when the brane disappears, 
a single confined flux tube will be recovered \cite{Sen}. 
{}From this point of view, ``confinement on the brane'' \cite{Yi} is
automatically realized: it directly follows from the confinement in
the bulk. 

The organization of this paper is as follows: in the next section, 
we consider one of such models in which the Abelian gauge field 
couples to a scalar field through the dielectric `constant'
\cite{dielectric}. It turns out that if we choose the potential 
and the dielectric constant appropriately, our model indeed has the
desired properties described above. Moreover, it is shown 
that the confined flux tube correctly becomes a source 
for the $B$-field in the bulk. This implies that the flux tube really 
represents a fundamental string. A relation to the tachyon
condensation in string theory is also addressed. 
In section 3, we propose other gauge theories which have the same
properties. In fact, for each confinement mechanism, it is possible 
to construct a gauge theory with the desired properties. 
This suggests that confinement in the bulk and
deconfinement on the brane is a quite universal phenomenon.  
Section 4 is devoted to discussions in which we put an emphasis 
on a possible relation to a fundamental string solution 
in a kind of IIB matrix model.

\section{Confinement via the Dielectric Effect}
\setcounter{equation}{0}

\subsection{The Model}

In this section, we consider the case in which the bulk electric flux
is confined by the dielectric effect proposed in \cite{dielectric}. 
Let us start with the Abelian gauge theory coupled to a scalar field
$\phi$ in $(d+1)$ dimensions considered in \cite{dielectric}
\be
{\cal L}=-\frac{1}{4}\vep(\phi)F_{\mn}^2
         +\frac{1}{2}\der^{\mu}\phi\der_{\mu}\phi-V(\phi)-j^{\mu}A_{\mu}
         +i\bar{\psi}\gamma^{\mu}\der_{\mu}\psi,
\label{model}
\ee
where 
\be
j^{\mu}=g\bar{\psi}\gamma^{\mu}\psi,
\ee
is the fermion current. Gauss' law tells us that 
\be
\der_{\mu}D^{\mn}=j^{\nu},
\label{Gausslaw}
\ee
where
\be 
D_{\mn}=\vep F_{\mn}.
\ee
The canonical momentum for $A_{\mu}$ is given by 
\be
\pi_A^{\mu}=-D^{0\mu}\equiv D^{\mu}.
\ee
In the following, we choose $A_0=0$ gauge. The Hamiltonian density 
is given by 
\be
{\cal H}=\frac{D^2}{2\vep}+{\cal H}_{\phi},
\ee
where we have assumed that there is no magnetic source and dropped 
the trivial fermion part. 

Let us assume that $D$ has a flux tube configuration in the $x^1$
direction. {}From Gauss' law (\ref{Gausslaw}), we obtain 
\be
f=\int dS~D_1,
\label{constraint}
\ee
where $S$ is a $(d-1)$-dimensional hyperplane orthogonal 
to the $x^1$-direction, and $f$ is the flux produced by the charge. 
Then we should examine whether there is a minimum of the tension 
(energy of the flux tube per unit length) under the constraint 
that a given total flux passes through the tube. 
Therefore, we minimize 
\be
T=\int dS\frac{D_1^2}{2\vep}-\la\left(\int dS~D_1-f\right),
\ee
where $\la$ is the Lagrange multiplier implementing the constraint 
(\ref{constraint}). It is easy to find that the solution satisfies 
\be
D_1=\la\vep.
\label{D}
\ee
{}From this solution, we find that the electric field $F_{01}$ 
is simply given by $\la$ and hence is constant. Nevertheless $D_1$ 
can be nontrivial due to the nontrivial dependence of the dielectric 
effect $\vep(\phi(x))$. Note that there is no flux in the region 
where $\vep$ is zero as seen from (\ref{D}). 

Solving the constraint by (\ref{D}), the energy per unit length 
can be rewritten as 
\be
W=\frac{1}{2}\frac{f^2}{\int\vep~dS}
 +\int dS\left( \frac{1}{2}((\vec{\nabla}\phi)^2+\dot{\phi}^2)
                    +V(\phi)
         \right),
\ee
where we have included the contribution from the scalar field. 
As shown in \cite{dielectric}, it is now easy to see that there exits 
a static configuration of $\phi$ which minimize $W$ under a suitable 
choice for $\vep$ and $V$. In \cite{dielectric}, they are given by 
\be
\vep=(\frac{\phi-\phi_0}{\phi_0})^4,
\label{dielectric}
\ee
\be
V(\phi)\sim\mu(\phi-\phi_0)^2.
\label{squarepot}
\ee
Thus, a nontrivial dependence of $\vep$ on $\phi$ allows 
a flux tube solution for $D$ which is energetically more favored 
than the spherically symmetric configuration. 
Therefore, if we put an electric charge in the bulk, it produces 
a flux tube which yields a linear potential between charges. 
This shows that this theory in the bulk is in the confinement phase. 

In our model, a new ingredient comes into the choice of the scalar potential
$V(\phi)$. As an example, the potential is chosen to be 
\be
V(\phi)=-\phi^2\log\phi^2,
\label{pot}
\ee
rather than (\ref{squarepot}) so that it can admit the unstable 
domain wall solution. In fact, this potential is studied in \cite{MZ} 
as a toy model of the tachyon condensation and is known to be 
the exact tachyon potential \cite{tachpot} in the context of 
the boundary string field theory (BSFT) \cite{BSFT}. 
As shown in \cite{MZ}, the unstable domain wall in this case is simply given 
as the Gaussian:
\be
\phi(x)=\exp(-x^2/4),
\label{gaussian}
\ee
where we have denoted the one-dimensional transverse coordinate 
of the domain wall (brane) as $x$. For illustration, let us take $\vep$ 
as 
\be
\vep(\phi)=\phi^4,
\label{dielectric2}
\ee
which is essentially the same as (\ref{dielectric}) up to an
irrelevant constant. In this case, $\vep(\phi)$ is zero in the bulk 
except in the flux tube produced by charges put there as described above.  
On the other hand, since $\vep$ is non-zero all over the unstable
$(d-1)$-brane, $D_1$ can be non-zero there. Therefore, it is natural 
to expect that the standard Abelian gauge theory is recovered on the
brane. In the next subsection, this claim is confirmed by examining 
the existence of the massless mode of the gauge field on the brane.

\subsection{Massless Mode on the Brane}

Let us examine the existence of the massless mode of the gauge field 
on the brane. The equation of motion for the gauge field far 
from a charge is 
\be
\der_{\mu}(\vep F^{\mn})=0. 
\label{Maxwell}
\ee
It is rather trivial that the $x$-independent mode of $A_{\mu}$ actually 
satisfies this equation as the massless mode. However, for completeness, 
let us solve this equation. 
Taking the $A_0=0$ gauge and assuming the plane wave solution for 
$A_x$ and $A_i$ with respect to $(t, x^i)$, where $x^i$ represents 
the longitudinal direction of the brane, we find that $\vep A_x$
satisfies the standard wave equation and that the equation of motion 
for $A_j$ is reduced to 
\be
\der_x^2A_i+\frac{\der_x\vep}{\vep}\der_x A_i+k_x^2A_i=0,
\ee
where $k_x$ is the momentum in the transverse direction of $A_x$. 
Substituting (\ref{gaussian}) and (\ref{dielectric2}) 
into this equation, we see that this equation is nothing but 
the Hermite differential equation and that the eigenvalue is given by 
\be
k_x^2=2n,~~~n=0,1,2,\cdots.
\ee
Thus the gauge field $A_i$ on the brane really has the massless mode 
and there is no tachyon even though the brane itself is unstable. 
The latter fact can be also seen directly from the equation of motion 
(\ref{Maxwell}).

\subsection{Coupling with the $B$-field}

In this subsection, we consider the effect of the bulk $B$-field. 
If the flux tube is really a fundamental string, it must be a source 
of the $B$-field in the bulk. For the purpose of including the
$B$-field into our Lagrangian, let us begin with the dual gauge theory 
in four dimension in which a magnetic flux is coupled to the bulk $B$-field:
\be
{\cal L}'=-\frac{1}{4}\vep^{-1}\tF_{\mn}^2
          -\frac{i}{2}\epsilon^{\mn\la\rho}\tF_{\mn}B_{\la\rho},
\ee
where $\tF$ is the dual field strength and, correspondingly, the
dielectric constant is inverted. Performing the duality transformation 
of this Lagrangian, we arrive at 
\be
{\cal L}=-\frac{1}{4}\vep(F_{\mn}-B_{\mn})^2, 
\ee
which seems quite natural. Therefore, we expect that in general the
coupling with $B$-field is introduced by replacing $F_{\mn}$ with 
$F_{\mn}-B_{\mn}$. Regarding the $B$-field as a background for the
gauge field and repeating the same argument as above, we obtain 
the Hamiltonian density as follows: 
\be
{\cal H}=\frac{D^2}{2\vep}+D^{\mu}B_{0\mu}.
\label{coupledtoB}
\ee
This shows that our flux tube correctly couples to the $B$-field. 
It is also easy to verify that $F_{0\mu}$ is again given by $\la$ 
and hence constant for the flux tube configuration. 
Notice that although $F_{0\mu}$ is constant for the flux tube 
configuration, $B$ is not necessarily constant. These facts suggest 
that the confined electric flux tube is a classical configuration 
in the gauge theory corresponding to the fundamental string. 
Therefore, if we omit the scalar field part, 
the complete Lagrangian must be 
\be
{\cal L}=-\frac{1}{4}\vep(F_{\mn}-B_{\mn})^2+cH_{\mn\rho}^2,
\ee
where $H$ is the field strength of the $B$-field, and we have not 
taken account of the effect by gravity. 

It is worth pointing out that the statements in this subsection 
are not restricted to the action of $F_{\mn}^2$ type. In fact, 
if we start from the `dual' Born-Infeld action 
\be
{\cal L}'=-\sqrt{\det(1+\tF)}
          -\frac{i}{2}\epsilon^{\mn\la\rho}\tF_{\mn}B_{\la\rho}
\ee
and performing the duality transformation developed in \cite{dualBI}, 
we obtain 
\be
{\cal L}=-\sqrt{\det(1+F-B)},
\ee
namely, the inclusion of $B$ amounts to making a replacement 
$F\rightarrow F-B$ as well in the Born-Infeld action. Furthermore, 
it can be shown that for a general Lagrangian 
${\cal L}((F_{\mn}-B_{\mn})^2)$, the electric field $F_{0\mu}$ 
becomes constant as above for the flux tube configuration.

\subsection{Stability of the Flux Tube}

In this section, we examine the stability of the flux tube 
against expanding it. For this purpose, let us examine how the total 
energy for the static flux tube 
\be
W=\frac{1}{2}\frac{f^2}{(\int\vep~dS)}
 +\int dS\left(\frac{1}{2}(\vec{\nabla}\phi)^2+V(\phi)
         \right),
\label{energy}
\ee
changes under a transformation 
\be
\phi(x^i) \rightarrow \phi(\la x^i),
\ee
where $x^i$ are the $d-1$ directions in which a section of $D$ has a 
support: $\int d^{d-1}x^i D=\int dS~D=f$. This transformation for $\la < 1$ 
corresponds to fattening the flux tube. More precisely, 
let us consider the transformation 
\be
\phi(x^i) \rightarrow \phi'(x^i)=\la^{\alpha}\phi(\la x^i),
\label{transf}
\ee 
where we choose $\alpha$ such that 
\be
\int \vep(\phi')~dS=\int \vep(\phi)~dS,
\label{epsiloninv}
\ee
namely, the above transformation decreases the range of $\phi$ 
simultaneously so that the integration of the dielectric term 
will keep invariant. 
Suppose $\vep(\phi) \sim \phi^m$, 
then a simple calculation shows that $\alpha=(d-1)/m$. 
If the potential behaves like $V(\phi) \sim \phi^n$, 
the kinetic term and potential term for the scalar field 
in (\ref{energy}) become under this transformation 
\bea
\int \frac{1}{2}(\vec{\nabla}\phi')^2 dS & 
     = & \la^{3-d+2\alpha}\int\frac{1}{2}(\vec{\nabla}\phi)^2 dS, \nono \\
\int V(\phi')~dS & = & \la^{1-d+\alpha n}\int V(\phi)~dS, 
\eea
where 
\bea
3-d+2\alpha & = & 3-d+\frac{2(d-1)}{m}, \nono \\
1-d+\alpha n & = & (d-1)\left(\frac{n}{m}-1\right).
\eea
Therefore, as long as $m$ is larger than the smaller value of $n$ 
and $2(d-1)/(d-3)$, expanding the soliton costs more energy 
of the gauge field. It is now evident that our choice (\ref{pot}), 
(\ref{dielectric2}) agrees with this condition. For $\la>1$, 
higher derivative terms possibly stabilize the flux tube. 
This establishes the stability of the fundamental string 
in the gauge theory (\ref{model}).

\subsection{Relation to the Tachyon Condensation} 

It is conjectured that when the open string tachyon around the unstable 
D-brane condenses, the gauge field on the brane forms 
a confined flux tube \cite{Yi} which plays a role of a piece 
of a closed string \cite{Sen}. It is further pointed out that 
in this case the tachyon potential plays a role of the dielectric 
constant \cite{ncsoliton,nothing}. In order to confirm this, 
let us apply the previous arguments to the Lagrangian 
\be
{\cal L}=-\frac{V(\phi)}{g_s}\sqrt{-\det(g+F)}
         -\frac{1}{g_s}\sqrt{-\det(g+F)}
          G_S^{\mn}\der_{\mu}\phi\der_{\nu}\phi,
\label{BI}
\ee
where $\phi$ is the tachyon field, $G_S^{\mn}$ is the symmetric part of 
$(g_{\mn}+F_{\mn})^{-1}$, and $V(\phi)$ is given as in (\ref{pot}). 
This Lagrangian is derived by using the boundary string field theory 
\cite{tachpot}.  

As before, we assume the flux tube solution along the $x^1$-direction 
and concentrate only on $F_{01}=E_1$. The Gauss' law constraint 
(\ref{constraint}) can be solved by 
\be
D_1=f\frac{\tV}{\int \tV dS}.
\label{BID}
\ee
and the minimized energy per unit length is given as 
\be
W=\sqrt{f^2+(\int\tV dS)^2}
 +\frac{\sqrt{f^2+(\int\tV dS)^2}}{\int\tV dS}
  \int (\der_{\mu}\phi)^2 dS,
\ee
where $\tV=V/g_s$. Around the tachyonic vacuum\footnote
{The phrase tachyonic vacuum refers to the vacuum after the tachyon 
condensates.}
, $\tV<<1$ and 
\be
W=f+\frac{(\int\tV dS)^2}{2f}
   +\frac{f}{\int\tV dS}\int (\der_{\mu}\phi)^2 dS.
\label{BIenergy}
\ee 
As seen from (\ref{BID}), the flux tube can exist only 
in the region where $\tV\neq 0$. In this sense, $\tV$ indeed plays 
a similar role to the dielectric constant near the tachyonic vacuum. 
Note that even if $\tV<<1$, $D_1$ can remain finite according to
(\ref{BID}). 
Thus we expect that the flux tube configuration satisfies both the minimum 
energy condition and the Gauss' law constraint. Moreover, as mentioned 
in the previous subsection, this flux tube correctly couples to 
the $B$-field. These facts seem to suggest that the flux tube is exactly 
the closed string at the tachyonic vacuum where the flux tube is
confined via the dielectric effect caused by the tachyon potential. 
However, the expression of $W$ implies that the flux tube in this case 
is unstable under the `fattening' transformation described in the 
previous subsection. In fact, it is easy to see that 
if we make the transformation (\ref{transf}) satisfying 
(\ref{epsiloninv}) with $\vep$ replaced by $\tV$, 
the first and second term in (\ref{BIenergy}) are invariant, 
while the last kinetic term decreases. Therefore, the configuration 
of $\phi$ spreads and eventually becomes flat. 
The flux tube is unstable. Since the tachyon potential (\ref{pot}) 
is known to be exact \cite{tachpot}, this fact suggests that 
the kinetic term should be modified if the flux tube really plays 
a role of the fundamental string at the tachyonic vacuum. 
Indeed, compared to the potential term, there is no good reason yet 
why the kinetic term can be still represented in terms of the open string 
metric as in (\ref{BI}) even near the tachyonic vacuum. In fact, 
the trouble in (\ref{BIenergy}) originates from the fact that 
$\tV$ plays both roles of the potential and of the dielectric constant. 
Thus one of the resolutions of this problem may be a modification 
in the kinetic term in (\ref{BI}).

\section{Other Models}
\setcounter{equation}{0}

In this section, we construct other gauge theories in which 
the fundamental string and the branes are realized as classical
configurations. The string is described by a confined flux tube
solution in the bulk and is deconfined on the branes. 
As origins of confinement other than the dielectric effect, 
we employ the non-Abelian gauge field and the vortex line. 
It turns out that for each confinement mechanism, it is possible 
to construct a gauge theory with this property.

\subsection{Confinement via the Non-Abelian Gauge Field}

Let us construct a four-dimensional non-Abelian gauge theory 
in which a confined flux tube in the bulk becomes deconfined 
on a domain wall of the scalar field coupled to the gauge field. 
For example, the Lagrangian is given 
by 
\be
{\cal L}=-\frac{1}{4g^2}F^a_{\mn}F^{a\mn}+|D_{\mu}\Phi|^2
         -(-\frac{v^2}{2}+\eta^2)|\Phi|^2-\frac{\kappa}{2}(|\Phi|^2)^2
         +\frac{1}{2}(\der_{\mu}\eta)^2-\la(\eta^2-v^2)^2,
\ee
where $F^a_{\mn}$ is the $SU(2)$ gauge field strength, 
$\Phi$ is a complex scalar field in the adjoint representation of
$SU(2)$, and $\eta$ is a real (neutral) scalar field. 
This model has already been considered in \cite{DS}. As shown in 
\cite{DS}, $\eta$ has a stable domain wall solution 
\be
\eta_0=v\tanh (\sqrt{2\la}vx). 
\label{kink}
\ee
In the bulk away from the domain wall, $\eta=\pm v$. In this case 
the potential for $\Phi$ has a positive mass term. Thus the bulk gauge field 
is the standard non-Abelian one and is in the confinement phase. 
Therefore, the string can be realized as a confined flux tube. 
On the other hand, at the core of the domain wall, 
$-v/\sqrt{2}<\eta<v/\sqrt{2}$, $\Phi$ has the negative mass 
term and consequently the $SU(2)$ gauge symmetry is spontaneously broken 
down to $U(1)$. Thus the flux is deconfined on the brane. 

In this example, we have a stable brane (\ref{kink}). It is also 
possible to construct an unstable brane solution by changing 
the form of the potential for $\eta$. For example, if we take an 
$\eta^3$ potential, there exists a lump solution. Then it is
possible to adjust parameters in such a way that 
$-v/\sqrt{2}<\eta<v/\sqrt{2}$ is satisfied only inside the
unstable brane.

\subsection{Confinement via the Vortex Line}

The second model is based on the Abelian Higgs model in which 
magnetic charges are confined by the vortex line proposed in
\cite{NO}. The Lagrangian is 
\be
{\cal L}=-\frac{1}{4}F_{\mn}F^{\mn}+|D_{\mu}\Phi|^2
         -(\frac{v^2}{2}-\eta^2)|\Phi|^2-\frac{\kappa}{2}(|\Phi|^2)^2
         +\frac{1}{2}(\der_{\mu}\eta)^2-\la(\eta^2-v^2)^2.
\ee
This model is different from the one in the last subsection in that 
$F_{\mn}$ is the Abelian gauge field and the sign in front of the mass 
term for $|\Phi|$ is opposite. In this case in the bulk the magnetic 
charges are confined via the vortex line. This flux tube can be
identified as a fundamental string in the bulk. Inside the brane 
the potential for $\Phi$ is the stable one, hence we have the 
standard Abelian gauge field in the Coulomb phase. 

In the dual picture, the electric flux is confined in the bulk 
and deconfined on the brane. This model is proposed in
\cite{Rubakov}. However, in this case the electric flux is 
dual to the magnetic one which is introduced by hand in order to 
cancel a singularity arising from a singular gauge transformation. 
In this sense, the electric flux tube is not a classical solution, 
but a kind of background. 

We conclude this section by making a remark on the confinement 
in the bulk and deconfinement on the brane. In the context of 
the tachyon condensation in open string theory, it is most likely 
that the bulk confinement is realized by the dielectric effect 
as described by the previous section. However, a variety of models 
in this section which realize the same situation suggest that 
this is a quite universal phenomenon.

\section{Discussions}
\setcounter{equation}{0}

In this section we discuss a relation between our model 
and the IIB matrix model. Motivated by (\ref{model}), let us 
consider a variant of the IIB matrix model 
\be
S=-\frac{1}{4}\Tr~\vep(Y)[A_{\mu},A_{\nu}]^2
  -\frac{1}{2}\Tr~(\bar{\psi}\Gamma^{\mu}[A_{\mu},\psi])
  +V(Y),
\label{NBIMM}
\ee 
where $A_{\mu}$ and $Y$ are bosonic $N\times N$ Hermitian matrices, 
and $\psi$ is a fermionic $N\times N$ matrix.  
$\vep(Y)$ and $V(Y)$ are assumed to be given as (\ref{dielectric2}) 
and (\ref{pot}) respectively. We see in (\ref{NBIMM}) that 
$\vep(Y)$ plays a role of the dielectric function 
and the dynamics of $Y$ is governed by the potential $V(Y)$. 
Of course the action is for the Yang-Mills field, but 
we may still expect that we have a confined Abelian flux tube 
in this model due to $\vep(Y)$. For example, as shown in \cite{NC}, 
if we expand (\ref{NBIMM}) around a following classical solution 
for $A_{\mu}$:
\be
[\hat{A}_{\mu}, \hat{A}_{\nu}]=iB_{\mn},
\label{NCback}
\ee
then the matrix model becomes the noncommutative $U(1)$ gauge theory 
with the dielectric function $\vep(Y)$. In this theory, $\vep(Y)$ is 
expected to confine a flux tube as well as in the commutative model 
considered in section 2. This confined tube should be also stable 
by the argument given in section 2.4. Thus we have a classical solution 
corresponding to a fundamental string in a kind of IIB matrix model. 
Electric and magnetic flux tube solutions in the noncommutative $U(1)$ 
gauge theories have been also obtained in
e.g. \cite{ncsoliton,ncsoliton2}, but the confinement problem has not 
been fully addressed.  

This type of model with $\vep(Y)\sim Y^{-1}$ and $V(Y)\sim Y$ 
has been proposed in \cite{NBI} as a nonperturbative regularization 
of the Schild action \cite{Schild} of type IIB superstring.
In their formulation, $Y$ is introduced to play the same role 
as $\sqrt{g}$ in the Schild action. It is also pointed out in
\cite{NBI} that $Y^{-1}$ can be regarded as the dielectric function 
proposed in \cite{dielectric}. In this sense, our model is quite 
similar to the one in \cite{NBI} although the potentials are different 
because of the different motivations. 

It must be emphasized that the above interpretation of a classical 
flux tube solution as the fundamental string in the IIB matrix model 
is conceptually different from the one in \cite{FKKT}. It would be
interesting to clarify their direct relationship.

\begin{center} \begin{large}
Acknowledgments
\end{large} \end{center}
We would like to thank S. Iso for useful discussions. 
The work of T.K. was supported in part by JSPS Research Fellowships 
for Young Scientists.

\end{document}